\documentclass[a4paper]{scrartcl}
\usepackage{graphics}
\usepackage{hyperref}
\def\address#1{}
\def\keywords#1{}
\usepackage[english]{babel}
\usepackage[latin1]{inputenc}
\usepackage[T1]{fontenc}
\usepackage{amssymb}
\usepackage{amsmath}
\usepackage{booktabs}

\def\ci#1{\includegraphics{r54_#1}} 

\let\epsilon=\varepsilon
\let\phi=\varphi

\def\op{{\oplus}}
\def\om{{\ominus}}

\def\rea{\rightarrow}

\def\wl{\overleftarrow{w}}
\def\wr{\overrightarrow{w}}

\begin{document}
\title{Gliders and Ether in Rule 54} \author{Markus Redeker}
\address{International Centre of Unconventional Computing, University
  of the West of England, Bristol, United Kingdom.
  Email: \email{markus2.redeker@live.uwe.ac.uk}}
\maketitle

\begin{abstract}
  This is a study of the one-dimensional elementary cellular automaton
  rule 54 in the new formalism of ``flexible time''. We derive
  algebraic expressions for groups of several cells and their
  evolution in time. With them we can describe the behaviour of simple
  periodic patterns like the ether and gliders in an efficient way. We
  use that to look into their behaviour in detail and find general
  formulas that characterise them.
\end{abstract}

\section{Introduction}

This is a case study of one specific cellular automaton, Rule 54, with
the methods developed in \cite{redeker2008}. They were developed to
allow the study of cellular automata with the methods of theoretical
mathematics and without the need for computer simulations. While the
previous paper concentrates on the development of the theory, here the
ideas are presented in a less formal way and used to work with larger
structures.

Section~\ref{sec:local-interactions} of this paper introduces the
formalism in a less formal way than in \cite{redeker2008} and shows
how the transition function of the cellular automaton can be expressed
in it. The resulting formulas still describe only the behaviour of a
small number of cells at a time. Therefore in
\begin{figure}[ht]
  \centering
  \ci{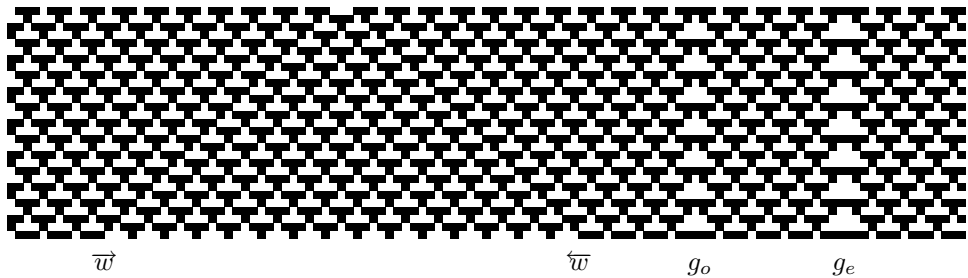}
  \caption{Periodic patterns under Rule 54. The diagram shows three
    types of gliders on an ether background. Time goes upward. (Since
    Rule 54 is symmetric, $\protect\wl$ and $\protect\wr$ are viewed
    as variants of the same particle.)}
  \label{fig:gliders}
\end{figure}
Section~\ref{sec:triangles} rules for larger groups of cells are
found. We use them in Section~\ref{sec:case-studies} to study the
behaviour of four simple periodic structures that occur under Rule 54:
the ether and three types of gliders (Figure~\ref{fig:gliders}). We
find formulas for them and general expressions for gliders and ethers
and look into their behaviour.

\section{Local Interactions}
\label{sec:local-interactions}

\subsection{Rule 54}

``Rule 54'' is the common name -- following the naming convention of
Stephen Wolfram \cite{wolfram1984} -- of a one-dimensional cellular
automaton with two states and a three-cell neighbourhood.

At every time it consists of an infinite line of cells. The
\emph{state} of each cell is an element of the set $\Sigma = \{0,
1\}$, and the behaviour of the automaton is given by its \emph{local
  transitions function}
\begin{equation}
  \phi: \Sigma^3 \longrightarrow \Sigma\,.
\end{equation}
It is applied to every three-cell subsequence of the infinite cell
line, and the resulting value is the state of the cell in the middle
at the next time step. Rule 54 has
\begin{equation}
  \label{eq:def54}
  \phi(s) =
  \left\{
    \begin{array}{r@{\quad}l}
      1&\mbox{for $s \in \{001, 100, 010, 101\}$,}\\
      0&\mbox{otherwise.}
    \end{array}
  \right.
\end{equation}
Sequences of elements of $\Sigma$ -- like $001$ -- stand here and
later for elements of $\Sigma^* = \bigcup_{k \geq 0} \Sigma^k$. Note
that $\phi$ is symmetric under the interchange of left and right.

\subsection{Situations}

The formalism of \emph{Flexible Time} \cite{redeker2008} is motivated
by the idea that it is easier to find patterns in the evolution of
cellular automata if one considers structures that involve cells at
different times.

These structures are here calles \emph{situations}. They are a
generalisation of the sequences of cell states (like 001) considered
before. These sequences give the states of neighbouring cells at a
certain unspecified time. Thus the sequence 001 describes the states
of three cells, possibly at the positions $x = 0$, $1$, $2$, and tells
us that the cells at $x = 0$ and $x = 1$ are in state 0 and the cell
at $x = 2$ is in state 1. The position information is implicit in the
ordering of the symbols: When a symbol in the sequence stands for the
state of a certain cell, its right neighbour in the sequence gives the
state of its right neighbour cell, and so on.

Situations are then cell sequences that extend over space and time. To
write them down we need additional symbols for a change of time. The
symbols we actually use stand for a displacement in time and also in
space, because they harmonise then better with the way a cellular
automaton evolves.

Under Rule 54, situations are written as sequences of the symbols 0,
1, $\om$ and $\op$. The intended interpretation can most easily be
described in terms of instructions to write symbols on a grid. The
fields of the grid are labelled by pairs $(t,x) \in \mathbb Z^2$; $x$
is the position of a cell and $t$ a time in its evolution. The writing
rules are then
\begin{itemize}
\item At the beginning the writing position is at $(0,0)$.
\item If the next symbol is 0 or 1, write it down and move the writing
  positions one step forward; if it was $(t, x)$ it is now $(t, x +
  1)$.
\item If the next symbol is $\om$, move the writing position from $(t,
  x)$ to $(t - 1, x - 1)$.
\item If the next symbol is $\op$, move the writing position from $(t,
  x)$ to $(t + 1, x - 1)$.
\item \emph{No overwriting}: One cannot write different symbols at the
  same field. (This concerns expressions like $01 \op\om 1$: After
  writing $01 \op\om$ one is again at position $(0,1)$ and tries to
  write a 1 in a field that contains already a 0. So $01 \op\om 1$ is
  not a valid situation, but $01 \op\om 0$ is.)
\end{itemize}
The result, in mathematical terms, is a function from a subset of
$\mathbb Z^2$ to $\Sigma$ together with an element of $\mathbb Z^2$
(the final writing position). The function, which is called $\pi_s$
for a situation $s$, describes the states of some cells at some times,
while the element of $\mathbb Z^2$, written $\delta(s)$, will be
important when parts of situations are substituted for others. The
whole situation is then the pair $s = (\pi_s, \delta(s))$. We will
also need an empty situation, which is written $\lambda$.

\begin{figure}[ht]
  \centering

  \def\arraystretch{4}
  \tabcolsep1.5em
  \begin{tabular}{lll}
    \ci{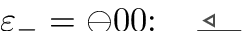} & \ci{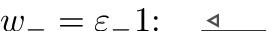} & \ci{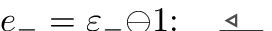} \\
    \ci{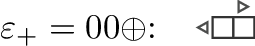}  & \ci{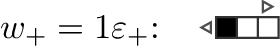}  & \ci{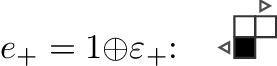} \\
  \end{tabular}

  \begin{tabular}{ll}
    \ci{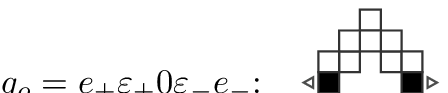} & \ci{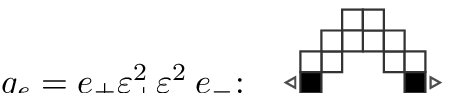} \\
  \end{tabular}
  \caption{Useful situations in Rule 54.}
  \label{fig:diagram54}
\end{figure}
In Figure~\ref{fig:diagram54} you can see diagrams for some situations
that will become useful later. Cells in the states 0 and 1 appear as
\ci{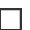} and \ci{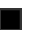}, while the initial and final writing
position are marked by small triangles: \ci{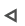} stands left of
the start position, \ci{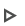} at the end position. The diagram
for $g_e$ becomes less surprising if one notices that
$\epsilon_-\epsilon_+ = 00 \op\om 00$ has the diagram \ci{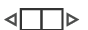}: a
first case of overwriting.

I have also treated there the situations as normal algebraic
expressions, like elements of a semigroup. Product and exponentiation
are defined in the usual way: $\epsilon^2$ is the result of writing
$\epsilon$ twice, and so on. However, due to the restrictions on
overwriting, not all products of situations exist.

\subsection{Reactions}

The evolution of cellular automata is described by \emph{reactions},
expressions of the form $a \rea b$ with two situations $a$ and $b$.
The situation $b$ represents a ``partially later'' state of the
cellular automaton than $a$, with the states of some cells at a later
time than $a$.

To make this notion more precise, let us consider functions of the
form $\pi: E \longrightarrow \Sigma$. They are called \emph{cellular
  processes} in \cite{redeker2008}. If a cellular process fulfills the
condition
\begin{equation}
  \label{eq:evolve}
  \begin{split}
    \text{If}\quad
    (t, x - 1), (t, x), (t, x + 1) \in E
    \quad&\text{then}\quad
    (t + 1, x) \in E\\
    &\text{and}\quad
    \pi(t + 1, x) = \phi(\pi(t, x-1)\pi(t, x)\pi(t, x+1)),
  \end{split}
\end{equation}
then it describes a part of the evolution of a cellular automaton
under the rule $\phi$.

With this notion we can define ``$\rea$'' as a binary relation on the
set of situations: $a \rea b$ is true if $\delta(a) = \delta(b)$ and
for all cellular processes $\pi$ that fulfill~\eqref{eq:evolve} we
have: If $\pi \supseteq \pi_a$ then $\pi \supseteq \pi_b$.

\begin{table}[t]
  \centering
  \def\r{$&${}\rea}
  \def\e{$&${}=}
  \tabcolsep=0pt
  \begin{tabular}{rl@{\qquad}rl}
    $\om 000  \r 0 \om 00$ & $000 \op \r 00 \op 0$ \\
    $\om 001  \r 1 \om 01$ & $100 \op \r 10 \op 1$ \\
    $\om 010  \r 11 \om 0$ & $010 \op \r 0 \op 11$ \\
    $\om 011  \r 00 \om 1$ & $110 \op \r 1 \op 00$ \\
    $\om 10   \r 1 \om 0 $ & $01 \op \r 0 \op 1$ \\
    $\om 11   \r 0 \om 1 $ & $11 \op \r 1 \op 0$ \\[1ex]

    $ 00 \e 00 \op\om 00$ & $ \om 00 \op   \r \lambda$ \\
    $ 01 \e 01 \op\om 01$ & $0 \om 1 \op 0 \r 0$  \\
    $ 10 \e 10 \op\om 10$ & $1 \om 1 \op 1 \r 1$ \\
    $ 11 \e 11 \op\om 11$ \\
  \end{tabular}
  \caption{Generator reactions for Rule 54}
  \label{tab:generator54}
\end{table}
One can see that if $xay$ and $xby$ are situations and there is a
reaction $a \rea b$, then $xay \rea xby$ is a reaction too. This is
called the \emph{application} of $a \rea b$ on $xay$. We can use that
and describe the behaviour of a cellular automaton by a small set of
\emph{generator reactions} between a small number of cells. All the
others follow from them by application on larger situations and by
chaining the reactions. Table~\ref{tab:generator54} shows a set of
generator reactions for Rule 54. It is derived from (\ref{eq:def54})
but contains some shortcuts.

To derive Table~\ref{tab:generator54} we start with the rule that
\begin{equation}
  \label{eq:phi-to-rea}
  \phi(\alpha\beta\gamma) = \sigma
  \qquad\mbox{becomes}\qquad
  \om \alpha\beta\gamma \rea \sigma\om\beta\gamma
  \qquad\mbox{and}\qquad
  \alpha\beta\gamma\op \rea \alpha\beta\op\sigma,
\end{equation}
because then $\sigma$ is placed correctly one time step later than
$\beta$. The first two lines of Table~\ref{tab:generator54} are found
this way. Other reactions, like $\om 10 \rea 1 \om 0$, are the result
of a unification: There would be both $\om 100 \rea 1 \om 00$ and $\om
101 \rea 1 \om 01$, but the state of the rightmost cell has no
influence on the result and is therefore left out at both sides of the
reaction. These new, shorter reactions can now be applied on the
results of some others: $\om 010 \rea 1 \om 10$, a reaction that one
gets from~(\ref{eq:phi-to-rea}), is then extended by $1 \om 10 \rea 11
\om 0$ to $\om 010 \rea 11 \om 0$. With these methods the top block of
Table~\ref{tab:generator54} is derived.

The purpose of the equations and reactions at the bottom of
Table~\ref{tab:generator54} is to create and destroy $\om$ and $\op$
symbols. The destruction reactions at the right remove also cell
states that cannot be used in another reaction.\footnote{In
  \cite{redeker2008}, which uses a slightly other definition of
  situations, the equations would have to be written as reactions. The
  destruction reactions, which are chosen somewhat ad hoc, are also
  different from the result of the result of the rules given there.}

Together the reactions of Table~\ref{tab:generator54} define a
\emph{reaction system} $\Phi$. It consists of a set of situations and
the reactions between them. We use a common convention and write $s
\in \Phi$ if $s$ is an element of the set of reactions of $\Phi$.

\section{A Reaction System with Triangles}
\label{sec:triangles}

\subsection{Triangles}

Now we need rules for larger structures. If their behaviour is
understood, we can find reaction that simulate them in one step. At
the present stage these structures will be periodic sequences of
cells, and we start with the simplest of then, the sequences in which
all cells are in the same state.

\begin{figure}[ht]
  \centering
  \tabcolsep2em
  \def\arraystretch{1.5}
  \begin{tabular}{cc}
    \ci{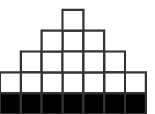} & \ci{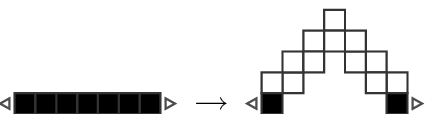}\\
    (a) Triangle &
    (b) Reaction $1^7 \rea 1 \op (0^2\op)^3 0 (\om 0^2)^3 \om 1$\\
  \end{tabular}
  \caption{Triangles and triangle reactions}
  \label{fig:triangles}
\end{figure}
There are only two of them, and we evolve them first for only one time
step.
\begin{alignat}{2}
  \label{eq:0step}
  0^k &\rea 0^2 \op 0^{k-2} \om 0^2 &\qquad k\geq 0\\
  \label{eq:1step}
  1^k &\rea 1 \op 0^k \om 1 &\qquad k\geq 0
\end{alignat}
We can see that $0^k$ is a \emph{persistent} pattern that reappears in
the next time step, while $1^k$ is \emph{instantaneous} and exists
only for one time step. Our guiding principle for a new, faster
reaction system will be that evolution should never stop when a
persistent pattern is reached.

So both reactions should be continued. The result for \eqref{eq:0step}
-- and therefore also for \eqref{eq:1step} -- depends on the parity of
$k$ and is best expressed as
\begin{alignat}{2}
  \label{eq:0triangle-raw}
  0^{2k+\iota} &\rea (0^2\op)^k 0^\iota (\om 0^2)^k
  &&\qquad k\geq 1, \iota \in \{0,1\}, \\
  \label{eq:1triangle-raw}
  1^{2k+\iota} &\rea
  1\op (0^2\op)^k 0^\iota (\om 0^2)^k \om 1
  &&\qquad k\geq 1, \iota \in \{0,1\}\,.
\end{alignat}
They are both examples of \emph{triangle reactions}, that are
reactions of the general form
\begin{equation}
  a_- x^k b_+ \rea a_+ y_+^k c y_-^k b_-
  \qquad k \geq 0,
\end{equation}
which trace the boundaries of a space-time triangle.\footnote{We can
  bring reaction~\eqref{eq:1triangle} in that form by setting $a_- =
  1^{2+\iota}$, $x = 1^2$, $a_+ = \lambda$, $y_+ = 0^2 \om$ $y_+ = \om
  0^2$ and $c = 0^\iota$.} Figure~\ref{fig:triangles} shows an
example.

Since the ``boundary terms'' of the triangles will occur often, we
will use abbreviations for them,
\begin{equation}
  \label{eq:boundary}
  \epsilon_- = \om 0^2,\qquad
  \epsilon_+ = 0^2 \op,\qquad
  e_- = \om 0^2 \om 1,\qquad
  e_+ = 1 \op 0^2 \op\,.
\end{equation}
The definitions for $e_-$ and $e_+$ have been chosen with the benefit
of hindsight -- instead of choosing abbreviations for $\om 1$ and $1
\op$ -- because these terms will be important later. With them
\eqref{eq:0triangle-raw} and \eqref{eq:1triangle-raw} become
\begin{alignat}{2}
  \label{eq:0triangle}
  0^{2k+\iota} &\rea \epsilon_+^k 0^\iota \epsilon_-^k
  &&\qquad k\geq 1, \iota \in \{0,1\}, \\
  \label{eq:1triangle}
  1^{2k+\iota} &\rea
  e_+ \epsilon_+^{k-1} 0^\iota \epsilon_-^{k-1} e_-
  &&\qquad k\geq 1, \iota \in \{0,1\}\,.
\end{alignat}

\subsection{Destruction of Boundary Terms}

\begin{table}[t]
  \centering
  \begin{tabular}{l@{\quad}l}
    \toprule
    States: & $0, 1, \epsilon_-$, $\epsilon_+$, $e_-$, $e_+$.\\
    \midrule
    Situations:
    & No subsequences $\epsilon_-0$, $0\epsilon_+$, $e_-1$, $1e_+$. \\
    \midrule
    Triangles: &
    \tabcolsep=0pt
    \begin{tabular}[t]{ll}
      $0^{2k+\iota} \rea \epsilon_+^k 0^\iota \epsilon_-^k$,
      &\qquad $k\geq 1$, $\iota \in \{0,1\}$ \\
      $1^{2k+\iota} \rea
      e_+ \epsilon_+^{k-1} 0^\iota \epsilon_-^{k-1} e_- $,
      &\qquad $k\geq 1$, $\iota \in \{0,1\}$
    \end{tabular} \\
    Boundary terms: &
    \tabcolsep=0pt
    \begin{tabular}[t]{rll}
      $\epsilon_- (10)^k 1 \epsilon_+$ & ${} \rea 1^{2k+3}$
      & \qquad $k\geq 0$ \\
      $\epsilon_- (10)^k e_+$ & ${} \rea 1^{2k+1} \epsilon_+$
      & \qquad $k\geq 0$ \\
      $e_- (01)^k \epsilon_+$ & ${} \rea \epsilon_- 1^{2k+1}$
      & \qquad $k\geq 0$ \\
      $e_- (01)^k 0 e_+$ & ${} \rea \epsilon_- 1^{2k+1} \epsilon_+$
      & \qquad $k\geq 0$ \\[0.5ex]

      $\epsilon_- \epsilon_+$ & ${}\rea \epsilon_+ \epsilon_-$ \\
      $e_- e_+$ & ${}\rea \epsilon_+ \epsilon_-$ \\
    \end{tabular} \\
    \bottomrule
  \end{tabular}
  \caption{Rule 54 in triangle form}
  \label{tab:triangle-form}
\end{table}
We must now extend these reactions to a full reaction system. Since
\eqref{eq:0triangle} and \eqref{eq:1triangle} create the boundary
terms $\epsilon_-$, $\epsilon_+$, $e_-$ and $e_+$, the new reactions
should destroy them. To keep the number of new reactions small, we
require that the triangle reactions are always used efficiently and
never applied to only a part of a cell sequence. (A reaction like $0^3
\rea \epsilon_+ \epsilon_- 0$ will be forbidden then.) We may express
that by the requirement that the situations may never contain the
terms $\epsilon_-0$, $0\epsilon_+$, $e_-1$ or $1e_+$: they would be
the result of such an incomplete application.

It will be enough for a working system to consider reactions that
start from terms of the form $b_- s b_+$, with $b_- \in \{\epsilon_-,
e_-\}$, $b_+ \in \{\epsilon_+, e_+\}$, $s \in \Sigma^*$, to which no
other reactions are applicable. The last requirement means that $s$
must consist of cells in states 0 and 1 in alternating order: Two
cells in the same state are already the starting point of a triangle
reaction. It turns out that there are only six types of reactions that
satisfy this requirement and that of the forbidden subconfigurations
in the previous paragraph.

Here they are, together with reactions that start from them:
\begin{alignat}{2}
  \epsilon_- (10)^k 1 \epsilon_+ & {} \rea 1^{2k+3}
  & \qquad k\geq 0, \\
  \epsilon_- (10)^k e_+ & {} \rea 1^{2k+1} \epsilon_+
  & \qquad k\geq 0, \\
  e_- (01)^k \epsilon_+ & {} \rea \epsilon_- 1^{2k+1}
  & \qquad k\geq 0, \\
  e_- (01)^k 0 e_+ & {} \rea \epsilon_- 1^{2k+1} \epsilon_+
  & \qquad k\geq 0, \\
  \label{eq:ether-epsilon}
  \epsilon_- \epsilon_+ & {}\rea \epsilon_+ \epsilon_-, \\
  e_- e_+ & {}\rea \epsilon_+ \epsilon_-\,.
\end{alignat}
The first four reactions have been chosen minimally such that the cell
states of $s$ in $b_-sb_+$ are replaced with states that are exactly
one time step later, such as in \eqref{eq:0step} and \eqref{eq:1step}.
The last two reactions cover the situations with $s = \lambda$ that
are not special cases of the previous four reactions. The resulting
reactions system is listed in Table~\ref{tab:triangle-form}.

\section{Ether and Gliders}
\label{sec:case-studies}

\subsection{The Ether}

\begin{table}[t]
  \centering
  \def\r{$&${}\rea}
  \tabcolsep=0pt
  \begin{tabular}[t]{rl@{\qquad}rl}
    $\epsilon_- e_+ \r 1 \epsilon_+$       & $\epsilon_- e_+ \r w_+$ \\
    $e_- \epsilon_+ \r \epsilon_- 1$       & $e_- \epsilon_+ \r w_-$ \\
    $\epsilon_- 1 \epsilon_+ \r e_+ 0 e_-$ & $w_- e_+ \r e_+ 0 e_-$ \\
                                         && $e_- w_+ \r e_+ 0 e_-$ \\
    $e_- 0 e_+ \r e+ 0 e_-$ \\
    $\epsilon_- 10 e_+ \r 1^3 \epsilon_+$  & $w_- 0 e_+ \r 1^3 \epsilon_+$\\
    $e_- 10 \epsilon_+ \r 1^3 \epsilon_+$  & $e_- 0 w_+ \r 1^3 \epsilon_+$\\
    $\epsilon_- 101 \epsilon_+ \r 1^5$     & $w_- w_+ \r 1^5$ \\
  \end{tabular}
  \caption{Simple Reactions that are useful in
    Section~\ref{sec:case-studies}. Most of them are special cases of
    Table~\ref{tab:triangle-form} or derived from them.}
  \label{tab:useful}
\end{table}
Now we will use the new reaction system to look at some phenomena that
occur under Rule 54. The first of them is the \emph{ether}, a robust
background pattern. It consists at alternating time steps of either the
cell sequence $01^3$ or $10^3$ infinitely repeated. (To verify the
reactions in this section Table~\ref{tab:useful} may be helpful.)

\begin{figure}[ht]
  \centering
  \def\arraystretch{1.1}
  \begin{tabular}{c}
    \ci{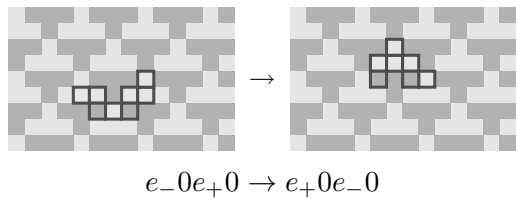}\\
    $e_- 0 e_+ 0 \rea e_+ 0 e_- 0$
  \end{tabular}
  \caption{How the ether reaction fits into the development if the
    ether. The cells that belong to the reaction are marked.}
  \label{fig:ether}
\end{figure}
In the reaction system a formula for the ether can be derived from the
$01^3$ generation: We have
\begin{equation}
  \label{eq:ethergen}
  01^3 \rea 0 e_+ 0 e_-
\end{equation}
and (see Figure~\ref{fig:ether})
\begin{equation}
  \label{eq:ether-e}
  0 e_- 0 e_+ \rea 0 e_+ 0 e_-,
\end{equation}
therefore
\begin{equation}
  (01^3)^k \rea (0 e_+)^k (0 e_-)^k
  \qquad k\geq 0,
\end{equation}
a very simple triangle reaction. This is in contrast to the other
possible starting point, $10^3$, where one gets
\begin{equation}
  \label{eq:ether-wrong}
  (10^3)^k \rea 1 \epsilon_+ (0 e_+)^k (0 e_-)^k \epsilon_- 1
  \qquad k\geq 1,
\end{equation}
a more complicated triangle reaction, in which also the components of
the other ether phase, $e_-$ and $e_+$, reappear. The reaction system
selects thus one of the phases of the ether as more natural than the
other, which is a helpful simplification.

If one now looks back at~\eqref{eq:ether-epsilon} and compares it
with~\eqref{eq:ether-e}, one sees that they follow a common pattern.
Both are \emph{background reactions} of the form
\begin{equation}
  \label{eq:background}
  b_- b_+ \rea b_+ b_-\,.
\end{equation}
This reaction can easily be iterated to $b_-^k b_+^\ell \rea b_+^\ell
b_-^k$, which describes the evolution of a large piece of a periodic
background pattern.

Their involvement in the ether is the reason why $e_-$ and $e_+$ got
their names in~\eqref{eq:boundary}.

\subsection{Gliders}

There are three kinds of long-lived structures that are described in
\cite{boccara1991} in some detail. There they are called particles,
now usually \emph{gliders}. There is one moving particle $w$, which
appears as $\wl$ and $\wr$, depending on the direction in which it
moves, and the ``odd'' and ``even gutter'' $g_0$ and $g_e$, which are
immobile.

\begin{figure}[ht]
  \centering
  \def\arraystretch{1.1}
  \begin{tabular}{c}
    \ci{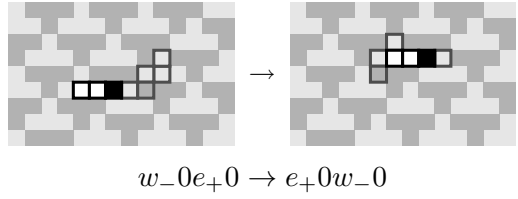}\\
    $w_- 0 e_+ 0 \rea e_+ 0 w_- 0$
  \end{tabular}
  \caption{A $\protect\wr$ glider moving on an ether background. The
    $w_-$ part is emphasised.}
  \label{fig:w-glider}
\end{figure}

The $w$ particle ``may be generated by three 0's followed by three 1's
or the converse'' \cite[p. 870]{boccara1991}. We try this now and get
\begin{equation}
  0^31^3 \rea \epsilon_+ 0 \epsilon_- e_+ 0 e_-
  \rea \epsilon_+ 0\; 1 \epsilon_+\; 0e_- .
\end{equation}
In it we can recognise $0e_-$ as a part of the ether and $\epsilon_+0$
as a part of the ether in the wrong phase (as in \eqref{eq:ether-e}
and \eqref{eq:ether-wrong}), so the rest must be the $w$ particle.
Therefore we define
\begin{equation}
  w_- = \epsilon_-1,\qquad
  w_+ = 1\epsilon_+.
\end{equation}
These definitions must be verified: We must show that $w$ actually
moves through the ether. But we have
\begin{equation}
  \label{eq:glider-w}
  w_- 0 e_+ 0
  = \epsilon_- 10 e_+ 0
  \rea 1^3 \epsilon_+ 0
  \rea e_+ 0 e_- \epsilon_+ 0
  \rea e_+ 0 \epsilon_- 1 0
  = e_+ 0 w_- 0,
\end{equation}
which shows how $w_-$ is destroyed and reappears at the right of its
previous position (Figure~\ref{fig:w-glider}). $w_-$ is therefore
stable and corresponds to the right-moving glider $\wr$ of
\cite{boccara1991}.

\begin{figure}[ht]
  \centering
  \def\arraystretch{1.5}
  \begin{tabular}{c}
    \ci{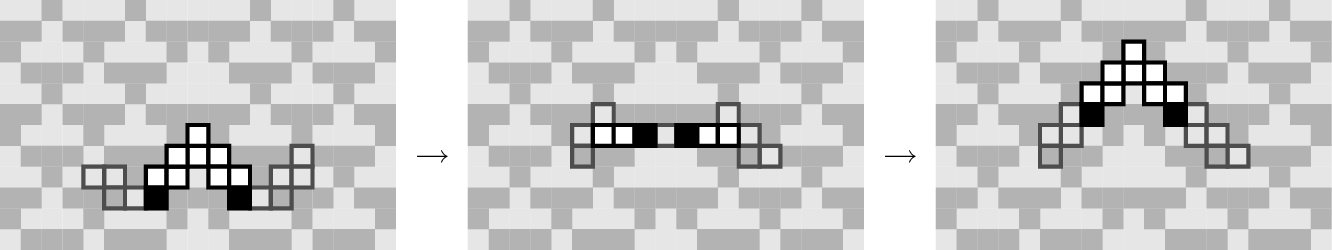}\\
    $e_- 0 g_o 0 e_+ 0 \rea e_+ 0 w_- 0 w_+ 0 e_- 0
    \rea e_+ 0 g_o 0 e_- 0$.\\[2ex]
    \ci{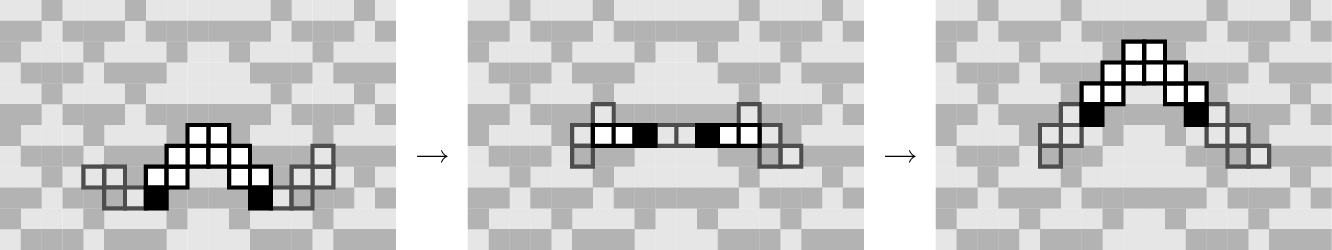}\\
    $e_- 0 g_e 0 e_+ 0 \rea e_+ 0 w_- \epsilon_-\epsilon_+ w_+ 0 e_- 0
    \rea e_+ 0 g_e 0 e_- 0$.
  \end{tabular}
  \caption{Evolution of the $g_o$ and $g_e$ gliders, together with the
    intermediate states where the $w$ gliders appear.}
  \label{fig:g-reaction}
\end{figure}

The two immobile gliders, $g_e$ and $g_o$, are in fact small
triangles, as can be seen from the pictures in \cite{boccara1991}. It
turns out that the right definitions for them are
\begin{equation}
  g_o = e_+ \epsilon_+ 0 \epsilon_- e_-,\qquad
  g_e = e_+ \epsilon_+^2 \epsilon_-^2 e_-.
\end{equation}
The verification that they actually behave like gliders is
straightforward (Figure \ref{fig:g-reaction}),
\begin{alignat}{2}
  e_- 0 g_o 0 e_+ 0 &
  = e_- 0 e_+ \epsilon_+ 0 \epsilon_- e_- 0 e_+ 0 \notag\\
  &\rea e_+ 0\; e_- \epsilon_+ 0 \epsilon_- e_+ 0 \;e_- 0 \notag\\
  &\rea e_+ 0\; w_- 0 w_+ 0 \;e_- 0 \notag\\
  &\rea e_+ 0\; 1^50 \;e_- 0 \notag\\
  &\rea e_+ 0\; e_+ \epsilon_+ 0 \epsilon_- e_- 0 \;e_- 0
  = e_+ 0 g_o 0 e_- 0, \label{eq:glider-go} \\
  e_- 0 g_e 0 e_+ 0 &
  = e_- 0 e_+ \epsilon_+^2 \epsilon_-^2 e_- 0 e_+ 0 \notag\\
  &\rea e_+ 0\; e_- \epsilon_+^2 \epsilon_-^2 e_+ 0 \;e_- 0 \notag\\
  &\rea e_+ 0\; w_- \epsilon_+ \epsilon_- w_+  0 \;e_- 0 \notag\\
  &\rea e_+ 0\; 1^6  0 \;e_- 0 \notag\\
  &\rea e_+ 0\; e_+ \epsilon_+^2 \epsilon_-^2 e_-  0 \;e_- 0
  = e_+ 0\; g_e  0 \;e_- 0, \label{eq:glider-ge}
\end{alignat}
but the appearance of the $w$ gliders in the process is a bit
surprising. It suggests the interpretation that the gliders $g_o$ and
$g_e$ decay into two $w$ gliders, which then collide and create its
next incarnation. With flexible time the gliders suddenly have an
internal structure.

The three glider reactions \eqref{eq:glider-w}, \eqref{eq:glider-go}
and \eqref{eq:glider-ge} have again a common structure, which can be
described by the \emph{glider condition}
\begin{equation}
  b_-^k g b_+^\ell \rea b_+^\ell g b_-^k.
\end{equation}
Here $b_-$ and $b_+$ form a background pattern as in
\eqref{eq:background} and $g$ is the glider. The number $(\ell - k) /
(\ell + k)$ is a measure for the speed of the glider.

We have now already touched the creation of other gliders by the $w$
gliders. Of the two syntheses found in the behaviour of the $g$
particles, the first one,
\begin{equation}
  w_- 0 w_+ \rea g_o,
\end{equation}
is more important because here the $w$ gliders are at the right
distance to have been part of the ether before. Such a glider
synthesis has been already noticed in \cite{boccara1991}, but here it
occurs as a corollary of a previous analysis.

\end{document}